\documentclass[a4paper]{jpconf}
\usepackage{graphicx}
\usepackage{physics}
\usepackage[bookmarks=false]{hyperref}
\usepackage[parfill]{parskip}	
\usepackage{amssymb}
\usepackage{color}
\usepackage{amsmath}
\usepackage{amsfonts}
\usepackage[square,sort&compress,numbers]{natbib}
\usepackage{doi}
\usepackage{subcaption}
\usepackage{dcolumn}
\usepackage{float}
\usepackage{lineno}
\usepackage[english]{babel}
\newcommand{\bo}{\beta_0}

\newcommand{\edb}{e^{-\Delta\beta E}}
\newcommand{\expbetaO}{\exp\qty(\beta_0 E)}
\newcommand{\mxpbetaO}{\exp\qty(-\beta_0 E)}
\newcommand{\zp}{\mathcal{Z}}
\begin{document}
\title{
    Using the Energy probability distribution zeros to obtain the critical properties of the two-dimensional anisotropic Heisenberg model}
\author{G.B.G. de Souza and B.V. Costa}
    \address{Laborat\'orio de Simulac\~ao, Departamento de F\'{\i}sica, ICEx Universidade Federal de Minas Gerais, 31720-901 Belo Horizonte, Minas Gerais, Brazil}
    \ead{ga.bruno926@gmail.com, bvc@fisica.ufmg.br}
\begin{abstract}
In this paper we present a Monte Carlo study of the critical behavior of the easy axis anisotropic Heisenberg spin model in two dimensions. Based on the partial knowledge of the zeros of the energy probability distribution we determine with good precision the phase diagram of the model obtaining the critical temperature and exponents for several values of the anisotropy. Our results indicate that the model is in the Ising universality class for any anisotropy.
\end{abstract}
\section{Introduction}
\label{Introd}
\noindent
    It is well known from rigorous results due to Mermin and Wagner and Hohenberg~\cite{mermin} that magnetization cannot exist in a Heisenberg magnet in two dimensions. However, the introduction of an exchange anisotropy multiplying one of the components of the spin, e.g., $S^{z}$, can induce a phase transition. Therefore we studied the model that is defined by the following hamiltonian
\begin{equation}
\label{Hamiltonian}
  H = -J \qty(\sum{S_{i}^{x}S_{j}^{x} + S_{i}^{y}S_{j}^{y} + \delta S_{i}^{z}S_{j}^{z}}) ~~ ,
\end{equation}
    where, $\vec{S}_{i}$ is a unit classical spin vector at a square lattice site $i$, the sum is extended over nearest-neighbor pairs, $J$ is the exchange constant, and $\delta$ is the anisotropy. This model has a continuum symmetry (considering the $xy$ plane) combined with a $Z(2)$ symmetry in the $z$-direction. Therefore, considering the limit as $T\rightarrow0$, if $\delta < 1$ the system has an infinite number of ground-states and the symmetry is not broken. Otherwise, if $\delta>1$, the ground state is two fold and the symmetry will be broken as the temperature increases. Since there is no frustration in the problem, the model described in (1) should and will be ruled by the Ising Field theory and the critical exponents must match the ones from the 2D Ising model for $\delta > 1$. However, Binder and Landau \cite{xxz_landau} studied this model for the case $\delta>1$ and their results supported the idea of a crossover, for an intermediate anisotropy, due to the reorientation of the magnetization from perpendicular to in-plane directions. Later, P.A. Serena \textit{et al.}~\cite{xxz_serena} studied the same model, using Monte Carlo. Due to the difficulties to properly equilibrate the system at low temperature they used a new algorithm exploring a restrict region of the phase space. Their results did not showed the crossover found earlier by Binder and Landau. They concluded the crossover was due to a lack of equilibration at low temperature. From our point of view, the Serena \textit{et al.} work can not be conclusive since the low temperature calculations could be biased due to the algorithm they used. Since then, as far as we know, there were no other related study on this model. The main goal in the present work is to study the Anisotropic Heisenberg Model (AHM) using the zeros of the Energy Probability Distribution (EPD) and a more sophisticated Monte Carlo approach as discussed in section \ref{Simulation}. The advantage of using the EPD method is that it does not demand the previous knowledge of any order parameter to get the critical temperature and the exponent $\nu$. We have obtained the critical temperatures and all critical exponents using finite size scaling for several values of the easy axis anisotropy. Our results for the critical temperature are consistent with previous works. Although they are expected to belong to the Ising universality class, as far as we know, there are no estimates for the critical exponents. It is noteworthy that our estimate for the critical temperature extends the results of reference~\cite{xxz_serena}. For each temperature we have obtained all critical exponents which does not indicates a possible crossover from the Ising class of universality, at large $\delta$, to another for small $\delta$.
    This paper is organized in the following way. In section \ref{Simulation} we discuss the simulation methodology used here, in particular, the Fisher zeros to settle a background to describe the EPD zeros approach. Next we describe the Monte Carlo method that combine three different algorithms together the "reweighting" technique. In Section \ref{Results} the numerical results are presented, including a finite size scaling analysis leading to the estimate of the critical temperatures and exponents. Finally, section \ref{Remarks} is devoted to our conclusions.
\section{Simulation Details}
\label{Simulation}
\noindent
\subsection{Fisher zeros}
\label{Fisher_}
\noindent
    A phase transition can be properly defined, with no ambiguity, by using the Fisher zeros~\cite{fisher1965nature}. Fisher has shown how the partition function can be written as a polynomial in terms of the variable $x = e^{-\beta \epsilon}$, where $\beta=1/{k_B T}$ is the inverse of the temperature, $k_B$ is the Boltzmann constant (Taken by simplicity as $k_B = 1$ from now on), and $\epsilon$ is the energy difference between two consecutive energy states of the system. The coefficients of the polynomial are real implying that their roots appear in conjugate pairs. If the system undergoes a phase transition at a certain temperature $T_c$, the corresponding zero, $x_c$, must be real and positive in the thermodynamic limit. For any finite system, all roots of the polynomial lie in the complex plane. In short it is as follows. The partition function of the system is given by
\begin{equation}
\label{z_part}
\mathcal{Z}_{N} = \sum_{E}g(E)e^ {-\beta E} =  e^{\beta \epsilon_{0}}\sum_{n=1}^{N}g_{n}\left( e^{\beta \epsilon}  \right)^{n} ~~~,
\end{equation}
    where it is assumed that the energy, $E$, can be written as a discrete set $E = \epsilon_0 + n\epsilon$ with $\epsilon_0$ some constant energy threshold. For a continuous energy distribution, like our model (Equation \ref{Hamiltonian}), a discretization of the density of states can be performed. If the phase transition occurs at $T_c$ the corresponding zero, $x_c(L)$, moves toward the positive real axis monotonically with the increasing size of the system as a power $L^{-1 / \nu}$. From now on we call it the \emph{dominant zero}. By plotting $\Im (x_c(L))$ and $\Re (x_c(L))$ as a function of $L$ the exponent $\nu$ and the critical temperature can be obtained. Although simple, this technique has some difficulties to be implemented due to the rapid grow in the number of energy states of the system. A suitable way to overcome this kind of problem is to use the EPD. It is closely related to the Fisher zeros as follows.
\subsection{Energy Probability Distribution zeros (EPD)}
\label{EPD_}
\noindent
     The EPD method to determine the dominant zero was developed in reference~\cite{epd_zeros,costa2019new,rodrigues2021moment}. It consists in judiciously selecting the most relevant zeros for a chosen temperature in the following way. Let us multiply equation \ref{z_part} by $1 = \expbetaO\mxpbetaO$, so that, it can be rewritten as
\begin{equation}
    \zp_{\bo} = \sum_E h_{\bo}(E) \edb  \qc \bigg\{
     \begin{array}{ll}
        h_{\bo} (E)  &= g(E) e^{\beta_0} ~~~, \\
        \Delta \beta  &= \beta - \beta_0 ~~~.
     \end{array}
\end{equation}
    Following the steps leading to equation \ref{z_part}, we define the variable $x = e^{-\Delta \beta \varepsilon}$ %
\begin{equation}
\label{zepd}
    \zp_{\bo} = e^{-\Delta\beta\varepsilon_0} \sum_n
      h_{\bo}(n) x^n ~~~,
\end{equation}
\noindent
    which is nothing but the \textit{energy probability distribution} of the canonical ensemble (without accounting for normalization), or simply the energy histogram at temperature $\bo$. Treating equation \ref{zepd} as a polynomial, there is an one to one correspondence between the Fisher zeros and the EPD zeros.
\begin{figure}[h]
  \centering
  \begin{minipage}[c]{0.5\textwidth}
    \includegraphics[width=\textwidth]{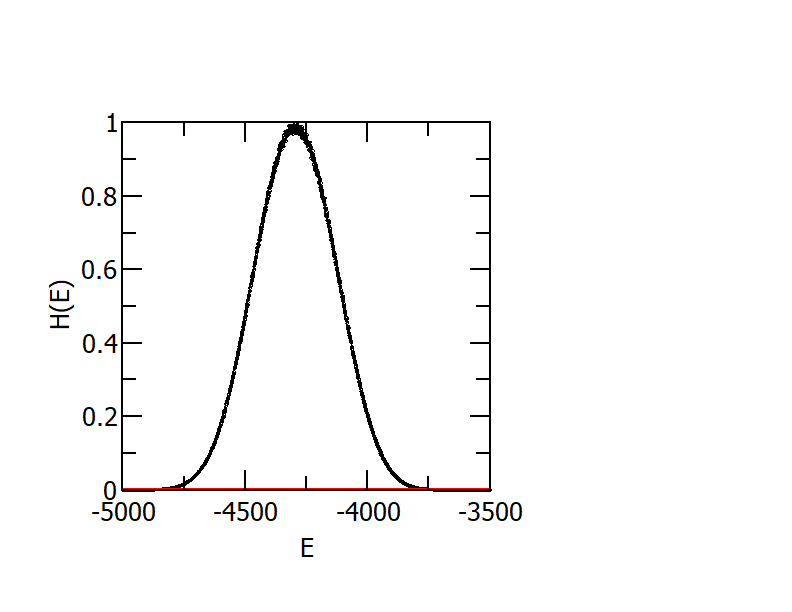}
  \end{minipage}
  \begin{minipage}[c]{0.5\textwidth}
    \includegraphics[width=\textwidth]{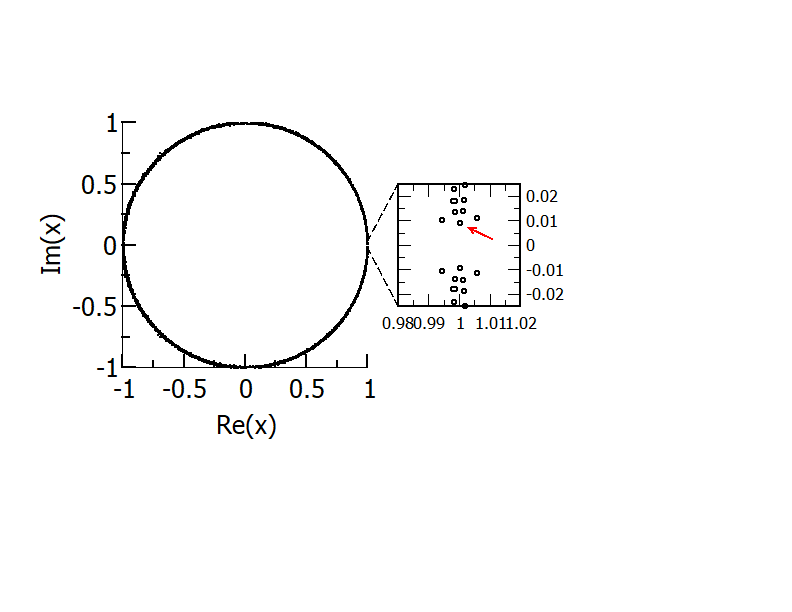}
  \end{minipage}
  \caption{
  (Color online) The figures show typical histogram (Top) and zeros distribution (Bottom). Here for $\delta = 1.50$ and $T = 1.3 J/k_B$. The red line in the histogram indicates the cutoff used. In the zeros distribution the zoomed figure shows the dominant zero, indicated with a red arrow.
  }
  \label{map_epd}
\end{figure}
\noindent
    For an histogram at the transition temperature, i.e., $\bo = \beta_c$, the dominant zero will be located over the real positive axis in the thermodynamic limit. For a finite but large enough system we expect a small imaginary part in $x_c(L)$. The dominant zero is the one with the smallest imaginary part regardless $\bo$. An important simplification can still be done. Only states with non-vanishing probability to occur are important to the phase transition. Therefore, we can discard \textit{small} values of $h_{\bo}$ by using some cutoff $h_{cut}$ (See figure \ref{map_epd}). Moreover, the dominant zero acts as an accumulation point, such that, even for initial histograms constructed far from the transition can give reasonable estimates for $\beta_c(L)$. Following this reasoning  a criterion to filter the important region in the energy space can be established. It goes as follows: First , build a normalized histogram $h_{\bo^0}$ (Max($h_{\bo^0}$)= 1) at an initial, but false, guess. Discard small values of $h_{\bo}$ according to $h_{cut}$ to reduce the polynomial degree. Then, construct the polynomial , Eq.\ref{zepd}, finding the corresponding zeros. By selecting the dominant zero, $x_c^0(L)$, we get an estimate for the pseudo-critical temperature $\beta_c^0(L)$. At the beginning of the process $\beta_c^0(L)$ is in general a crude estimate of $\beta_c(L)$. Nevertheless, we can proceed iteratively making $\beta_0^1(L) = \beta_c^0(L)$, building a new histogram at this temperature and starting over. After a reasonable number of iterations we may expect that $\beta_c^j(L)$ converges to the true $\beta_c(L)$ and thus $x_c^j(L)$ approaches the point (1,0). This process corresponds to applying a sequence of transformations, $\mathfrak{M}$, such that we reach the fixed point $\beta^{n+1}(L) = \mathfrak{M} \beta^n(L)$.
\begin{figure}[h]
  \centering
  \begin{subfigure}[b]{0.4\textwidth}
    \includegraphics[width=\textwidth]{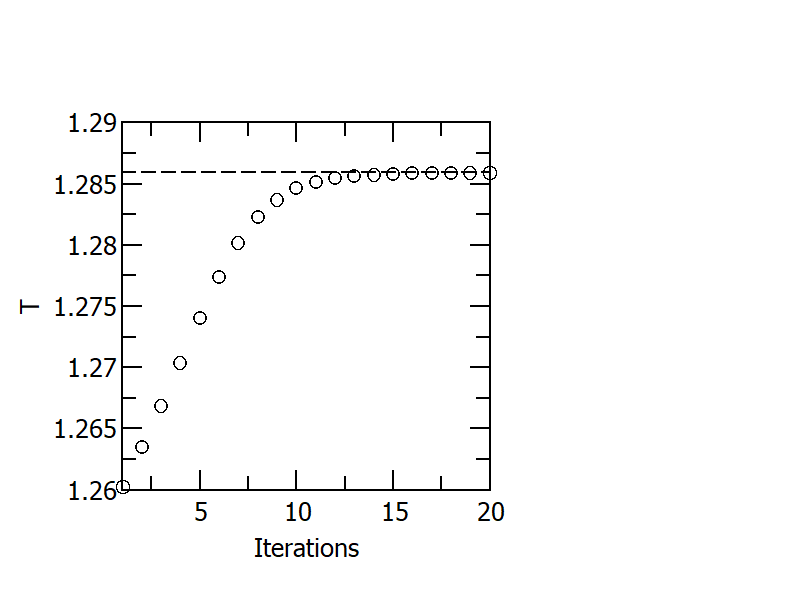}
\caption{}
  \end{subfigure}
  \hfill
  \begin{subfigure}[b]{0.4\textwidth}
    \includegraphics[width=\textwidth]{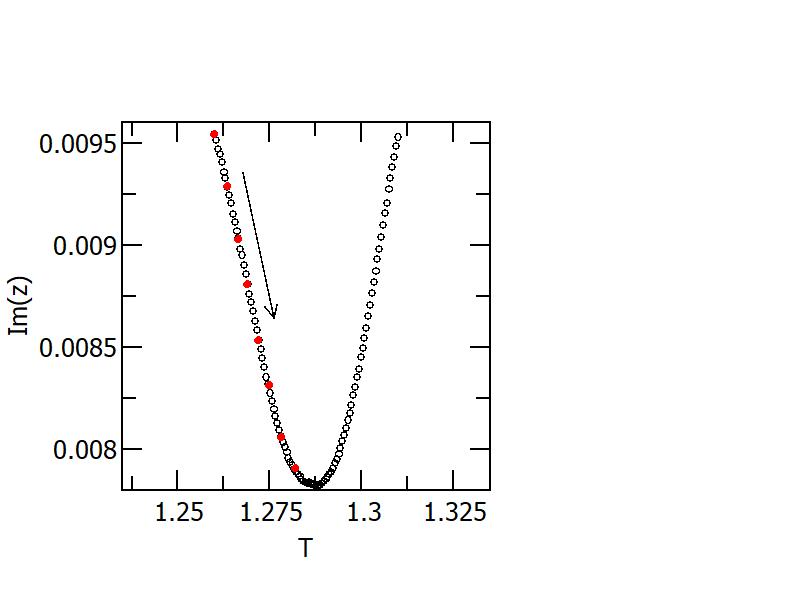}\caption{}
  \end{subfigure}
  \caption{
  (Color online) Example of the evolution of the iterative process to determine the dominant zero for $\delta = 1.5$ and $L = 60$ leading to $T_c(L) \approx 1.29$. The figure in the lrs shows the temperature for several iterations converging to the dominant zero (dotted-dashed line). The rhs figure shows the geometrical structure with an attraction basin in the vicinity of the dominant zero. The red dots show the evolution of the dominant zero when using the \emph{Regula Falsi} method, as described in the text.
  }
  \label{conv_epd}
\end{figure}
    This iterative process is illustrated in figure \ref{conv_epd} for $ \delta = 1.5$ and $L=60$. The open circles were obtained by selecting temperatures to show the geometrical structure near the dominant zero. The solid red circles are for the iterative process. The existence of an attraction basin can be clearly seen. Next we discuss the process in obtaining the histograms.
\subsection{Numerical Details}
\label{Simulation_tech}
\noindent
    In our simulation each Monte Carlo step (MCS) consists of $4$ single spin Metropolis update~\cite{metropolis} combined with one Overrelaxation move~\cite{super} and $1$ cluster Wolff updates~\cite{wolff} over the entire lattice. First we sweep all the $L^2$ spins in the lattice proposing a Metropolis move for each spin. After performing a Metropolis sweep four times, we make one over-relaxation move for each spin followed by one Wolff cluster update. It is worth noticing that, considering the symmetry of the hamiltonian \ref{Hamiltonian}, the Wolff and the Over-relaxation must be done only in the planar components. The first $100\times L^2$ MCS are discarded in order to reach equilibration. After thermalization we used $2\times 10^6$ MCS storing energy $E$, $E^2$, out-of-plane magnetization $M_z$ and $M_z^2$ at each MCS to build the histograms.
\subsection{Histograms}
\label{Histograms}
\noindent
    The histograms were taken as a post-simulation procedure, i.e., we run the simulations to initially obtain the raw data. Working in this way is important as it avoids defining the discretization as a simulation parameter and allow us to find the most appropriate discretization without having to rerun the simulations. In this work we used $\dd E = 0.5 J$. To build the single histograms we followed a "reweighting" recipe by Ferrenberg and Swendsen~\cite{mhist} to extrapolate the results for temperatures in the vicinity of the simulated ones. The cutoff we used was $h_{cut} = 10^{-3}$. To get the zeros we used the software \emph{MPSolve}~\cite{mpsolve1,mpsolve2}. Once the dominant zero is located the thermodynamic quantities as energy $\expval{E}$, specific heat $C_v = \beta^2 [\mathrm{ Var}(E) ]$, out-of-plane magnetization $\expval{M_z}$ and out-of-plane susceptibility $\chi_z = \beta [\mathrm{ Var}(M_z) ] $ are calculated in the vicinity of the dominant zero using the reweighting technique. Typical results are shown in figure \ref{Termo_Quantities}) for $\delta = 1.5$. Each point in our calculations is the result of an average over five independent histograms.
\begin{figure}[h!]
  \centering
  \begin{subfigure}[b]{0.4\linewidth}
    \includegraphics[width=\linewidth]{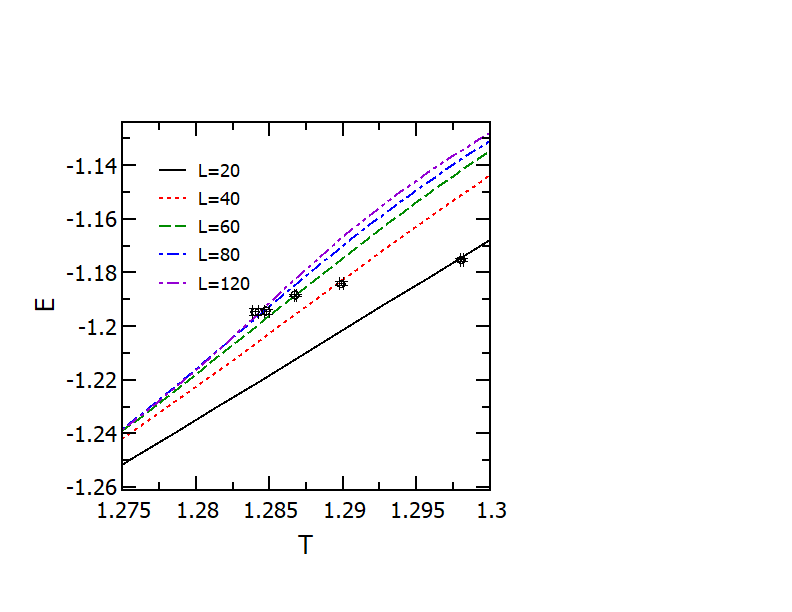}
\caption{}
  \end{subfigure}
  \begin{subfigure}[b]{0.4\linewidth}
    \includegraphics[width=\linewidth]{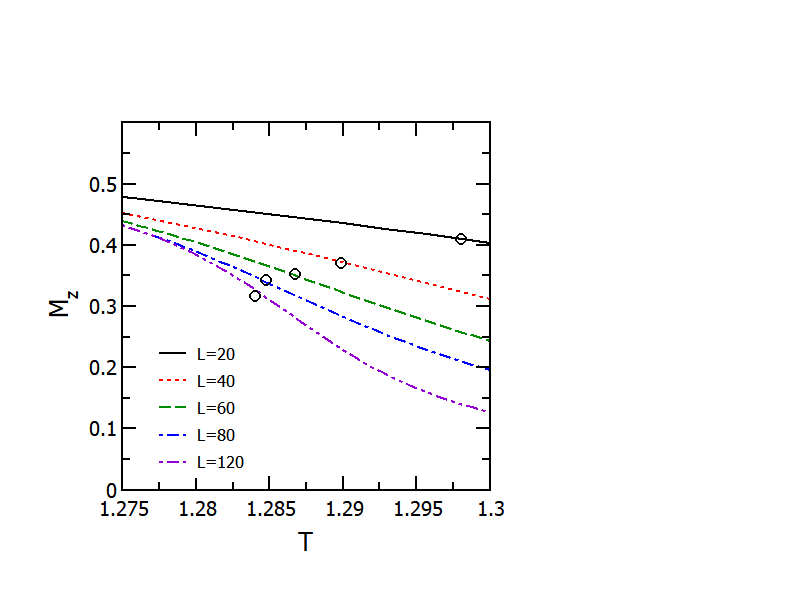}
\caption{}
  \end{subfigure}
  \begin{subfigure}[b]{0.4\linewidth}
    \includegraphics[width=\linewidth]{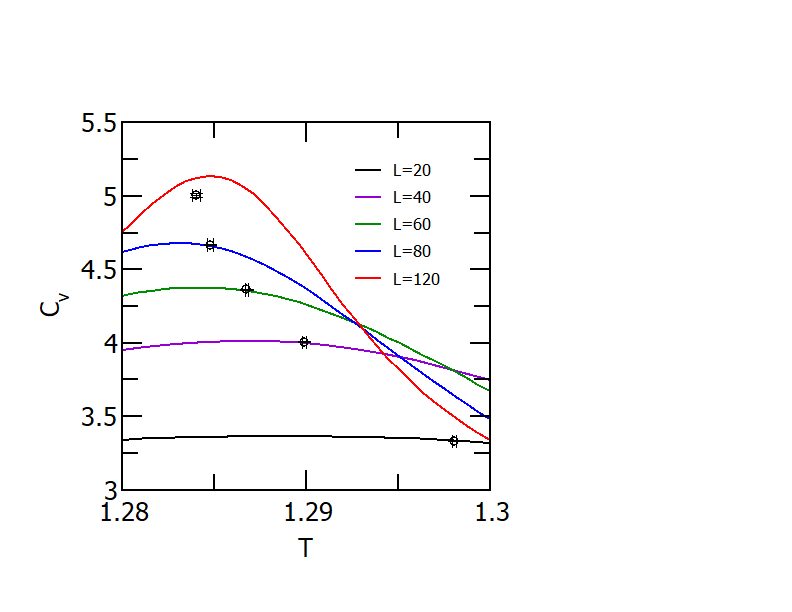}
\caption{}
  \end{subfigure}
  \begin{subfigure}[b]{0.4\linewidth}
    \includegraphics[width=\linewidth]{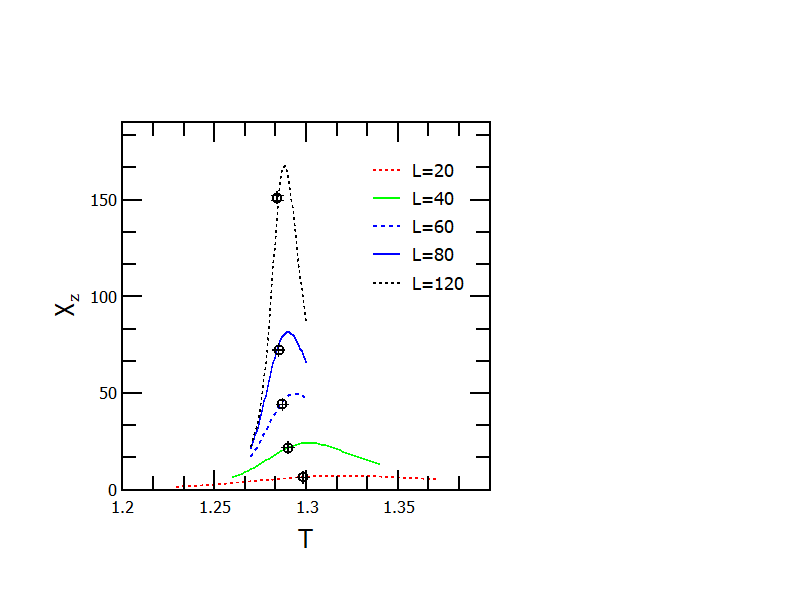}
\caption{}
  \end{subfigure}
    \caption{(Color Online) Energy, magnetization, specific heat and the susceptibility in (a), (b), (c) and (d) respectively, as a function of temperature for $\delta = 1.5$ for several lattice sizes, $L$. The highlighted points in each figure are the pseudo-critical temperatures obtained using the EPD technique as discussed in the text. Error bars are smaller then the symbols when not shown.}
\end{figure}
    \label{Termo_Quantities}
\noindent
\subsection{Finite-Size Scaling}
\label{FSS}
\noindent
    The finite-size scaling~\cite{privman, barkema_fss} theory provides us a way to extrapolate the results obtained from finite systems to the thermodynamic limit. The basic assumption of FSS is that on the vicinity of a phase transition the thermodynamic quantities are homogeneous functions of their arguments and should depend only on the ratio between the relevant dimension $L$ of the system and the correlation length, $\xi$, in such way they behave as~\cite{privman1990finite,barkema_fss}
\begin{align}
    \label{cv_l}
    C_v &\approx L^{\alpha/\nu} \mathcal{C}(tL^{1/\nu})\\
    \label{chi_l}
    \chi_z &\approx L^{\gamma/\nu} \mathcal{X}(tL^{1/\nu}) \\
    \label{Mz_l}
    M_z &\approx L^{-\beta/\nu} \mathcal{M}(tL^{1/\nu})
\end{align}
    where $\mathcal{C},\mathcal{X},\mathcal{M}$ are the proper derivatives of the free energy and $t \equiv \frac{T - T_c}{T_c}$. At $t = 0$ that quantities do not depend on $L$ so that, $C_v, \chi_z$ and $M_z$, follows a pure power law. A similar expression can be written for the pseudo-critical temperature $T_c(L)$~\cite{ITZYKSON1983415}
\begin{align}
\label{tc_l}
    T_c(L) \approx T_c + \lambda L^{-1/\nu} ~~~,  \\
\label{im_l}
\imaginary{m[x_c]}(L) \approx y_0 L^{-1/\nu} ~~~.
\end{align}
\noindent
    Although we do not have a rigorous demonstration of equation \ref{im_l}, numerical works  strongly suggests it holds~\cite{epd_zeros,costa2019new,rodrigues2021moment}. Using the FSS equations above, the critical temperature and exponents can be easily obtained from the simulation data. With $\nu$ in hand equation \ref{tc_l} readily furnishes the critical temperature. The hyper-scaling laws~\cite{privman1990finite,barkema_fss}, $d\nu = 2 - \alpha$ and $\alpha + 2\beta + \gamma = 2$ can be used as checking conditions.
\section{Results}
\label{Results}
\begin{figure}[h]
  \centering
  \begin{subfigure}[b]{0.4\linewidth}
    \includegraphics[width=\linewidth]{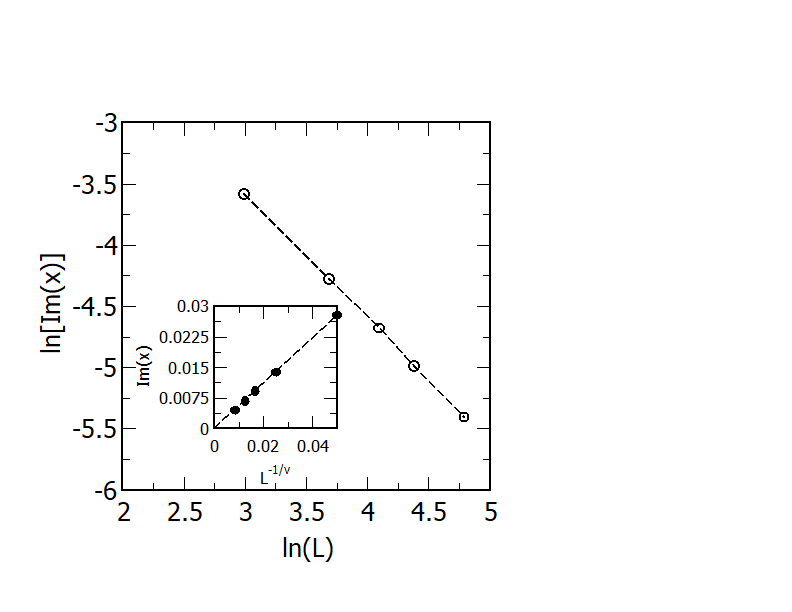}
\caption{}
  \end{subfigure}
  \begin{subfigure}[b]{0.4\linewidth}
    \includegraphics[width=\linewidth]{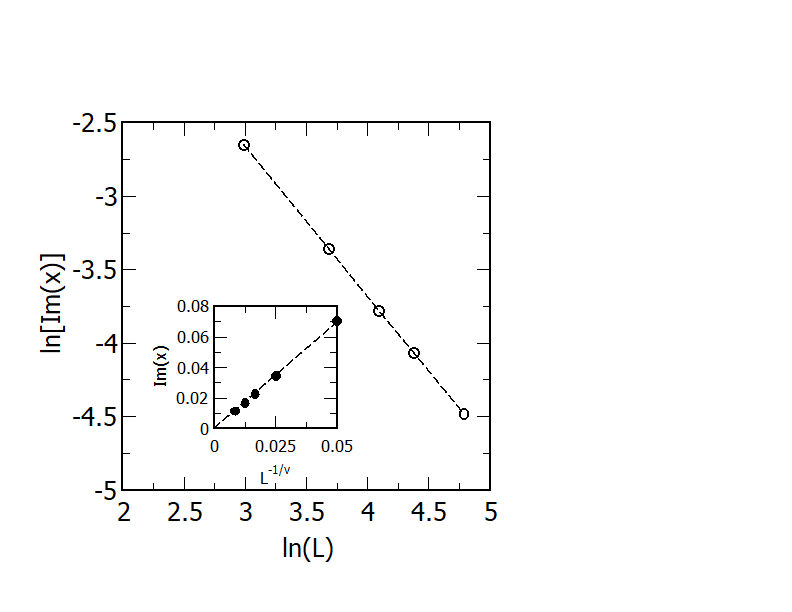}
\caption{}
  \end{subfigure}
    \caption{(Color Online) Typical finite size scaling analysis for the imaginary part of $z(L)$. Here we used $\delta = 1.5$ and $\delta = 1.05$ respectively. Error bars are smaller then the symbols when not shown.}
    \label{FSS_z}
\end{figure}

\begin{figure}[h!]
  \centering
  \begin{subfigure}[b]{0.4\linewidth}
    \includegraphics[width=\linewidth]{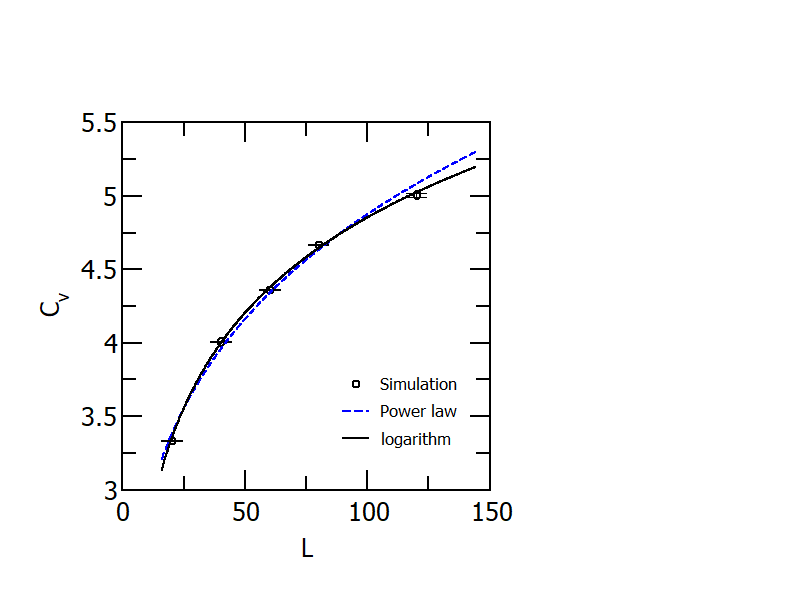}
\caption{}
  \end{subfigure}
  \begin{subfigure}[b]{0.4\linewidth}
    \includegraphics[width=\linewidth]{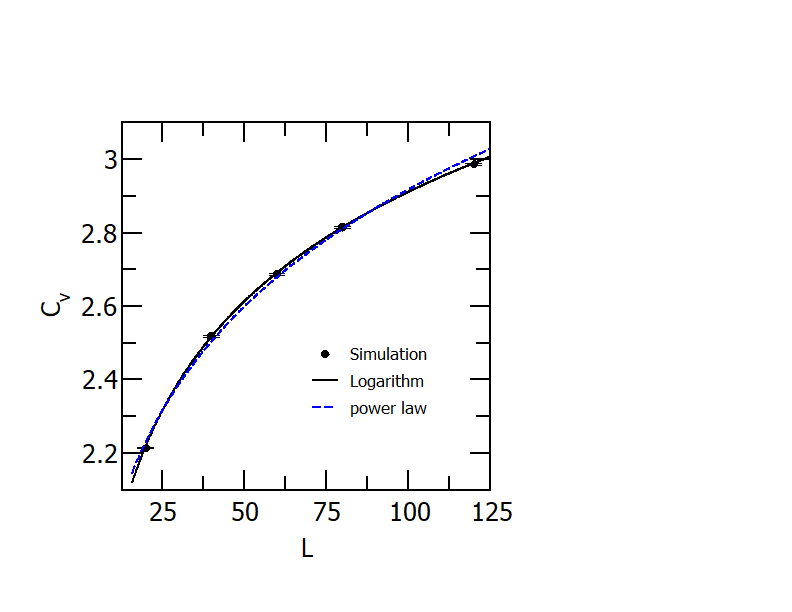}
\caption{}
  \end{subfigure}
    \caption{(Color Online) The figures (a) and (b) are typical results for the maxima of the specific heat, here, for $\delta = 1.5$ and $\delta = 1.05$ respectively. The points are from simulation data. The dashed line corresponds to a logarithm adjust and the dotted-dashed line is a power-law adjust. Clearly the logarithm behavior fits better the simulation data which is confirmed by a $\chi^2$ analysis. Error bars are smaller then the symbols when not shown.}
    \label{alpha}
\end{figure}

\noindent
    Initially, we use the imaginary part of the dominant zero to obtain the critical exponent $\nu$. As $Im(x) \sim L^{-1/\nu}$ a linear adjust of $ln[Im(x)] \times ln(L)$ furnishes $\nu$. Typical results are shown in figures \ref{FSS_z} for $\delta = 1.50$ and $1.05$. The insets in the figures are for $Im(x) \times L$. Once the exponent $\nu$ is obtained the critical temperatures can be estimated by adjusting $T_c(L) \times L^{-1/\nu}$.    To obtain the remaining exponents, $\alpha, \beta$ and $\gamma$ we use the values of the corresponding quantities at the pseudo-critical temperature. In particular, we draw attention to the determination of the $\alpha$ exponent, shown for two typical cases ($\delta = 1.50$ and $1.05$) in figure \ref{alpha} for $C_v(L) \times L$. Clearly, a $C_v(L) \times ln(L)$ fits the data better than $C_c(L) \times L^{\alpha/\nu}$. We assume that $\alpha = 0$ for all $\delta$.
    Our results are summarized in tables \ref{Tab:1}. As a matter of comparison the data from Serena et al.~\cite{xxz_serena} for the critical temperatures are included in table \ref{Tab:1} and  figure~\ref{phase_diagram} for the phase diagram. In table \ref{Tab:2} are shown the deviations from the hyper-scaling relations.
\noindent
\begin{center}
\begin{table}[t]
\caption{The table shows the results obtained from the simulation. The entries with $()^{*}$ are from reference \protect\cite{xxz_serena}.}
\label{Tab:1}
\begin{tabular}{|c|c|c|c|c|c|c|}
\hline
  $\delta$        & $\beta$  & $\gamma$ & $\nu$    &  $T_c$ \\
\hline
\hline
  1.500           & 0.136(7) & 1.76(3)  & 0.990(5) & 1.2811(3) \\
  \hline
  1.200           & 0.129(4) & 1.71(2)  & 0.996(6) & 0.9543(4) \\
  \hline
  1.100           & 0.123(5) & 1.69(3)  & 0.982(9) & 0.8320(2) \\
  \hline
  $1.111^*$       &          &          &          & 0.76(1)   \\
  \hline
  1.050           & 0.120(9) & 1.65(6)  & 0.979(3) &  0.7432(4) \\
  \hline
  $1.010^*$       &          &          &          & 0.66(12)   \\
  \hline
  $1.001^*$       &          &          &          & 0.59(15)   \\
  \hline
  \textbf{Ising}  & 0.125    &   1.75   &     1    &            \\
  \hline
\end{tabular}
\end{table}
\noindent
\begin{figure}
    \includegraphics[width=\linewidth]{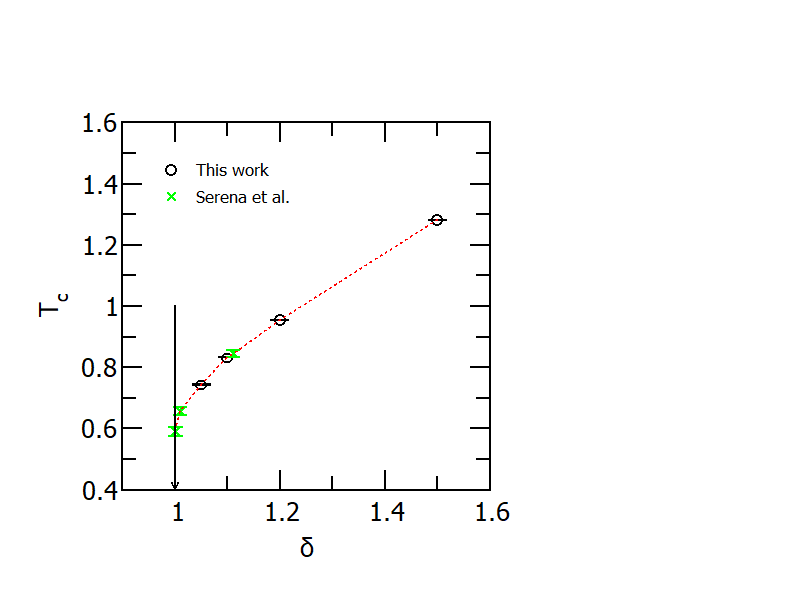}
    \caption{(Color Online) This figure compares the critical temperatures we obtained (circle dots) with those present in P.A. Serena \textit{et al.} (square). The solid line shows the isotropic Heisenberg limit, $\delta = 1$, where $T_c = 0$. }
        \label{phase_diagram}
\end{figure}
\end{center}
\begin{table}[t]
\caption{This table shows how much the simulation data deviates from the hyper-scaling relations: $\alpha+2\beta+\gamma-2=0$ and $d\nu -2 = 0$. The exponent $\alpha$ is always assumed as $\alpha = 0$ as discussed in the text. The last line shows the exact Ising critical exponents.}
\label{Tab:2}
\begin{tabular}{|c|c|c|c|c|}
\hline
$\delta$  & $2 - 2\beta-\gamma = 0$ & $d\nu - 2 = 0$  \\
\hline
\hline
  1.500   & 0.030(5)              & 0.020(2)     \\
  \hline
  1.200   & 0.032(8)              & 0.032(2)     \\
  \hline
  1.100   & 0.064(5)              & 0.036(3)     \\
  \hline
  1.050   & 0.071(5)              & 0.042(20)     \\
  \hline
\end{tabular}
\end{table}
\noindent
\newpage
      The critical exponents clearly indicates an Ising-like behavior~\cite{stanley1987introduction} although there are some variations of the numerical values when compared with the ones from the Ising Model. Perhaps the crossover claimed by Serena \textit{et al} is due to finite size effects around $\delta=1$ and not necessarily due to lack of equilibration at low temperature.
\newpage
\section{Final Remarks}
\label{Remarks}
\noindent
    The main results in this work can be summarized as follows: (1) The study of the critical properties of a system, using the energy probability distribution (EPD) technique is robust, giving reliable results even when applied to moderate size systems. (2) For the anisotropic Heisenberg model in two dimensions, the results indicate the model is in the Ising universality class for any easy-axis anisotropy, $\delta > 1$. The variations of the values presented in Table 1 from the values of the Ising model are probably due to imprecision of the lattice sizes. (3) We have determined the critical temperature up to $5$ figures, extending earlier results.
    The determination of the critical exponents is always a difficult task. In general, the determination of $\nu$ and $T_c$ has to be done simultaneously, however, using the EPD technique, $\nu$ is obtained independently. As it is obtained from the imaginary part of the dominant zero  we have the advantage of knowing in advance that $\imaginary{m[x_c]}(L) \rightarrow 0$ and $L^{-1/{\nu}} \rightarrow 0$ in the thermodynamic limit that, can serve as a check for the confidence of the simulation.
%
\ack
    This work was partially supported by CNPq and Fapemig, Brazilian Agencies. GB and BVC thanks CNPq and FAPEMIG for the support under grants CNPq 402091/2012-4, CNPq 130064/2021-1 and FAPEMIG RED-00458-16.
\section*{Declarations}
\textbf{Conflict of Interests} The authors have no conflict of interests to declare. The financial support to this manuscript have already been declared on the "Acknowledgments". We certify that tho submission is original work and is not under review as any other publication.
\newcommand{\newblock}{}
\bibliographystyle{iopart-num}

\begin{thebibliography}{}
    \bibitem{mermin}Mermin, N. \& Wagner, H. Absence of Ferromagnetism  or Antiferromagnetism in One- or Two-Dimensional Isotropic  Heisenberg Models. {\em Phys. Rev. Lett.}. \textbf{17}, 1133-1136 (1966,11)
\bibitem{xxz_landau}Binder, K. \& Landau, D. Critical properties of the two-dimensional anisotropic Heisenberg model. {\em Phys. Rev. B}. \textbf{13}, 1140-1155 (1976,2)
\bibitem{epd_zeros}Costa, B., Mól, L. \& Rocha, J. Energy probability distribution zeros: A route to study phase transitions. {\em Computer Physics Communications}. \textbf{216} pp. 77-83 (2017)
\bibitem{costa2019new}Costa, B., Mól, L. \& Rocha, J. A new algorithm to study the critical behavior of topological phase transitions. {\em Brazilian Journal Of Physics}. \textbf{49}, 271-276 (2019)
\bibitem{rodrigues2021moment}Rodrigues, R., Costa, B. \& Mól, L. Moment-generating function zeros in the study of phase transitions. {\em Physical Review E}. \textbf{104}, 064103 (2021)
\bibitem{fisher1965nature}Fisher, M. The Nature of Critical Points. (University of Colorado Press,1965)
\bibitem{privman}Privman, V. Finite Size Scaling and Numerical Simulation of Statistical Systems. (World Scientific,1990)
\bibitem{metropolis}Metropolis, N., Rosenbluth, A., Rosenbluth, M., Teller, A. \& Teller, E. Equation of State Calculations by Fast Computing Machines. {\em The Journal Of Chemical Physics}. \textbf{21}, 1087-1092 (1953)
\bibitem{wolff}Wolff, U. Collective Monte Carlo Updating for Spin Systems. {\em Phys. Rev. Lett.}. \textbf{62}, 361-364 (1989,1)
\bibitem{super}Creutz, M. Overrelaxation and Monte Carlo simulation. {\em Phys. Rev. D}. \textbf{36}, 515-519 (1987,7)
\bibitem{mhist}Ferrenberg, A. \& Swendsen, R. Optimized Monte Carlo data analysis. {\em Phys. Rev. Lett.}. \textbf{63}, 1195-1198 (1989,9)
\bibitem{barkema_fss}Newman, M. \& Barkema, G. Monte Carlo Methods in Statistical Physics. (Clarendon Press,1999)
\bibitem{privman1990finite}Privman, V. Finite size scaling and numerical simulation of statistical systems. (World Scientific,1990)
\bibitem{ITZYKSON1983415}Itzykson, C., Pearson, R. \& Zuber, J. Distribution of zeros in Ising and gauge models. {\em Nuclear Physics B}. \textbf{220}, 415-433 (1983)
\bibitem{stanley1987introduction}Stanley, H. Introduction to Phase Transitions and Critical Phenomena. (Oxford University Press,1987)
\bibitem{mpsolve1}Bini, Dario A., Fiorentino, Giuseppe, Design, analysis, and implementation of a multiprecision polynomial rootfinder. Numerical Algorithms 23.2-3 (2000): 127-173.
\bibitem{mpsolve2}Bini, Dario A., and Robol, Leonardo. Solving secular and polynomial equations: A multiprecision algorithm. Journal of Computational and Applied Mathematics 272 (2014): 276-292.
\bibitem{xxz_serena}Serena, P., Garcia, N. \& Levanyuk, A. Monte Carlo calculations on the two-dimensional anisotropic Heisenberg model. {\em Phys. Rev. B}. \textbf{47}, 5027-5036 (1993,3)
\end{thebibliography}

\end{document}